\documentclass[a4paper,11pt]{article}

\usepackage[latin1]{inputenc}
\usepackage[american]{babel}
\usepackage[dvips]{graphicx}
\usepackage{amsmath} \usepackage{amsfonts} \usepackage{amssymb}
\usepackage{bbm}
\DeclareMathOperator{\sgn}{Sgn}

\begin{document}

\newcommand{\partfrac}[2]{\frac{\partial #1}{\partial #2}}
\newcommand{\re}{\,\mathbb{R}\mbox{e}\,}
\newcommand{\im}{\,\mathbb{I}\mbox{m}\,}
\newcommand{\hyph}[1]{$#1$\nobreakdash-\hspace{0pt}}
\newcommand{\dvol}{d\mbox{Vol}_6}
\newcommand{\vol}{\mbox{Vol}_6}
\newcommand{\cir}{\mathaccent23}
\providecommand{\abs}[1]{\lvert#1\rvert}

\newcommand{\Nugual}[1]{$\mathcal{N}= #1 $}

\numberwithin{equation}{section}

\begin{flushright}
IC/2006/067\\
SISSA-45/2006/EP

\end{flushright}
\begin{flushright}
\end{flushright}
\begin{center}
\huge{ \sc Fixing Moduli in Exact Type IIA Flux Vacua}\\
\vskip 2cm

\Large{\sc B.S. Acharya$^a$, F. Benini$^b$ and R. Valandro$^b$ 
}
\\
\vskip 15mm \normalsize
{\sf $^a$ Abdus Salam ICTP,
Strada Costiera 11,\\ 34014 Trieste, ITALIA\\
\smallskip
{\it bacharya@ictp.it}\\
\vskip 5mm
$^b$ International School for Advanced Studies (SISSA/ISAS),\\Via Beirut 2-4, 34014 Trieste, ITALIA \\
\smallskip
{\it benini@sissa.it}\\{\it valandro@sissa.it}}

\end{center}
\vskip 15mm
\begin{center}
{\bf {\sc Abstract}}
\end{center}

Type IIA flux compactifications with $O6$-planes have been argued from a four dimensional
effective theory point of view to admit stable, moduli free solutions. We discuss in detail the
ten dimensional description of such vacua and present exact solutions in the case when the
$O6$-charge is smoothly distributed. In the localised case, the solution is a half-flat, non-Calabi-Yau
metric. Finally, using the ten dimensional description we show how all moduli are stabilised 
and reproduce precisely the results of de Wolfe et al.~ \cite{deWolfe}.

\newpage

\normalsize

\section{Introduction} \label{Intro}

String vacua with magnetic fields along the extra dimensions (`flux compactifications') have been intensively
studied in recent years (see \cite{Grana} for a recent review). One reason for their relevance is that, since the flux contribution to the energy depends on the geometrical moduli of the internal manifold, it gives them a four dimensional effective potential and can thus stabilize some or all of them, lifting undesired massless fields \cite{polchstrom,becker2,michelson,dasguptarajeshsethi,gkp,BsA,KKLT}.

Type IIA flux vacua are perhaps the best understood amongst flux vacua (see \cite{deWolfe,Kach,GrimL,VilZ,CamFI} and references therein). This is because
all the moduli are stabilised classically i.e. the effective moduli potential generated by the
tree level supergravity action in ten dimensions (supplemented with orientifold 6-plane sources)
has stable isolated critical points. This has been demonstrated in detail in \cite{deWolfe}.

Specifically, if we consider Type IIA theory on a Calabi-Yau threefold, switching on the RR fluxes
gives rise to a potential which depends on the K\"ahler moduli. In order to stabilise the complex 
structure moduli one can introduce NSNS 3-form flux, H, however this leads to a tadpole for the
\hyph{D6}brane charge, which can be cancelled by introducing orientifold six-planes ($O6$). The full system of fluxes and
\hyph{O6}planes then stabilises all the moduli, essentially at leading order in $\alpha'$ and $g_s$.

In particular, de Wolfe et al. \cite{deWolfe} have described the effective 4d potential for the moduli in the large volume
limit, when the backreaction of the fluxes on Einstein's equations can be ignored (since their contribution
to the stress tensor is volume suppressed). This class of vacua is an excellent arena to study aspects of moduli
stabilisation in detail, since the vacua are essentially classical solutions of ten dimensional IIA supergravity. However, until now, very little is known about what these ten dimensional solutions look like, since most of the prior studies have used the effective four dimensional description. 
The purpose of this paper is thus to fill this gap.

The basic questions we will ask are: does the ten dimensional solution actually exist (i.e. is the
four dimensional description valid)? If so, what, {\it precisely}, is the backreaction of the fluxes and
how does it modify the Ricci flat Calabi-Yau metric? Can we understand moduli stabilisation from a ten dimensional perspective?

Our main results can be summarised as follows: we prove that the exact ten dimensional solution is
{\it not} Calabi-Yau. The precise modification of the Calabi-Yau geometry can be described by a particular
type of {\it half-flat} SU(3) structure \cite{salamon}. Notably, they appear in the mirror-symmetric picture of `Calabi-Yau with fluxes' compactifications \cite{Gurrieri:2002wz, Gurrieri:2002iw}. Though we were unable to find the full solution (for which we will have to await further developments in the mathematical literature), in the
approximation that the \hyph{O6}plane source is smoothed out, we found an exact solution. This solution is
Calabi-Yau and by studing the moduli stabilization from the ten dimensional point of view, we found the same results as \cite{deWolfe}.

\

The paper is organized as follows. In section 2 we shortly review a class of solutions of Type IIA supergravity found in \cite{BehrCev} and \cite{Lust}.  These will form the basis of the solutions with \hyph{O6}planes.
They describe compactifications on an internal $SU(3)$-structure manifold down to four dimensional $AdS_4$. In section 3 we discuss the introduction of orientifold 6-planes in supergravity, the issue of supersymmetry preserving configurations and how the original solutions are modified by their presence. In particular, we present an exact ``smeared'' solution in which the orientifold charge is smoothed out. Finally in section 4 moduli stabilization is studied. We find that all the geometrical moduli are lifted at tree level in supersymmetric vacua. Conventions, supersymmetry variations and $SU(3)$-structure relations are relegated to some appendices.

We would also like to mention that Banks and van den Broek have also been studying similar issues to those
discussed here \cite{banks}.

\section{Massive Type IIA Supergravity on $AdS_4$} \label{LustSol}

Recently, a large class of supersymmetric four dimensional smooth compactifications of massive Type IIA supergravity have been classified \cite{Lust}.
In this section we will briefly review these solutions in order to set the notation for our results.
Following this, we will describe how the solutions are modified when \hyph{O6}planes are added.

The massive IIA theory has bosonic fields consisting of a metric $g$, an RR 1 form potential $A$ (with field strength $F$)
and 3-form potential $C$ (with field strength $G$), a NSNS 2-form potential
$B$ (with field strength $H$) and a dilaton $\phi$.

\medskip

We are interested in the ten dimensional description of the supersymmetric vacua with non-zero cosmological constant discussed by de Wolfe et al
from an effective field theory point of view in \cite{deWolfe}. Therefore, without loss of generality, we can take the ten dimensional
spacetime to be a warped product $AdS_4 \times_{\Delta} X_6$, where $X_6$ is a compact manifold and the ten dimensional metric is given by
\begin{equation} \label{eq021}
g_{MN}(x,y)=\left(
  \begin{array}{cc}
    \Delta^2(y)\hat{g}_{\mu\nu}(x) & 0\\ 0& g_{mn}(y)\\
  \end{array}\right) \:,
\end{equation}
where $x$ and $y$ are coordinates for $AdS_4$ and $X_6$ respectively and the warp factor is $\Delta$.
All the fluxes have non-zero $y$-dependent components only along the compact directions, except for
$G$ which has a non-zero four-dimensional component
\begin{equation}
G_{\mu\nu\rho\sigma} = \sqrt{g_4} f(y) \epsilon_{\mu\nu\rho\sigma} \:,
\end{equation}
and $f$ is a function on $X_6$. These assumptions are dictated by local Poincar\'e invariance on $AdS_4$.

$\mathcal{N}=1$ SUSY in four dimensions implies that the compact manifold $X_6$ has a globally defined spinor, $\eta$.
As a consequence, the structure group of $X_6$ reduces (at least) to $SU(3)$.
As usual, the existence of the spinor $\eta$ implies the existence of a globally defined 2-form $J$ and 3-form $\Omega$:
\begin{gather}\label{eq005}
J_{mn} \equiv i \eta_{-}^\dagger \gamma_{mn}\eta_{-} = -i \eta_{+}^\dagger \gamma_{mn}\eta_{+} \\
\label{eq005a}
\Omega_{mnp} \equiv \eta_{-}^\dagger \gamma_{mnp}\eta_{+} \qquad
\Omega^*_{mnp} = -\eta_{+}^\dagger \gamma_{mnp}\eta_{-} \:,
\end{gather}

With these properties $J$ and $\Omega$ completely specify an \hyph{SU(3)}structure on $X_6$. $J$ defines an almost complex structure with respect to which $\Omega$ is $(3,0)$.
From the $SU(3)$ decomposition of their differentials $dJ$ and $d\Omega$, one can read off the torsion classes
which characterize the \hyph{SU(3)}structure:
\begin{equation}\label{eq008}
\begin{split}
dJ &= -\frac{3}{2} \mbox{Im}(\mathcal{W}_1\Omega^*) +\mathcal{W}_4\wedge J +\mathcal{W}_3 \\
d\Omega &= \mathcal{W}_1 J\wedge J + \mathcal{W}_2\wedge J +  \mathcal{W}_5^*\wedge\Omega
\end{split} \end{equation}

\

By requiring the fluxes to preserve precisely $\mathcal{N}=1$ SUSY in four dimensions,
the ten dimensional supersymmetry parameter has to be of the form \cite{Grana}:
\begin{align} \label{eq004}
\begin{split}
\epsilon &= \epsilon_+ + \epsilon_- \\
     &= (\alpha \theta_+ \otimes \eta_+ - \alpha^* \theta_- \otimes \eta_-)+
        (\beta \theta_+ \otimes \eta_- - \beta^* \theta_- \otimes \eta_+) \:.
\end{split}
\end{align}
Here $\theta_+$ and $\theta_-$ (with
$\bar \theta_+= \theta_-^T C$) are the two Weyl spinors on $AdS_4$, satisfying the Killing spinor equations
\begin{equation}
\hat{\nabla}_\mu \theta_+ = W \hat{\gamma}_\mu \theta_- \qquad
\hat{\nabla}_\mu \theta_- = W^* \hat{\gamma}_\mu \theta_+  \:,
\end{equation}
where $W$ is related to the scalar curvature $\hat{R}$ of $AdS_4$ through $\hat{R}=-24\abs{W}^2$. On the other hand, $\eta_+$ and $\eta_-$ are chiral spinors on $X_6$ related by charge conjugation, so that $\epsilon$ is a Majorana spinor.

\

By substituting this ansatz in the SUSY equations $\delta \Psi_M =0$, $\delta \lambda=0$, Lust and Tsimpis find the following
solutions:\footnote{The 10 dimensional action, the supersymetry variations and the conventions are set in the appendix \ref{action}.}

If  $\abs{\alpha}\not = \abs{\beta}$, one gets the usual Calabi-Yau supersymmetric
compactification, i.e. $X_6$ is a Calabi-Yau manifold, all the fluxes vanish and $W=0$, so the four dimensional space is Minkowski.

If $\abs{\alpha} = \abs{\beta}$, one can, without loss of generality, choose $\alpha = \beta$ and:
\begin{equation} \label{eq006}
\begin{split}
F &= \frac{f}{9} e^{-\phi/2} J + \tilde{F} \\
H   &= \frac{4m}{5} e^{7\phi/4} \re \Omega \\
G   &= f d\mbox{Vol}_4 + \frac{3m}{5} e^{\phi} J\wedge J  \\
W   &= \Delta \left(\frac{\alpha}{\abs{\alpha}}\right)^{-2} (-\frac{1}{5}m \, e^{5\phi/4} + \frac{i}{6}f \, e^{\phi/4}) \\
&\phi,\Delta,f, \mbox{Arg}(\alpha) = \mbox{constant} \:.
\end{split}
\end{equation}
Here $\tilde F$ is the $\mathbf{8}$ component in the $SU(3)$ decomposition of $F$ (see appendix \ref{action}) and it
is not determined by supersymmetry. On the other hand, by imposing the Bianchi identities, one finds a contraint on its differential:
\begin{equation}
d\tilde{F} = - \frac{2}{27}e^{-\phi/4} \left( f^2 - \frac{108}{5} m^2 \, e^{2\phi} \right) \re\Omega \:.
\end{equation}
From the last equation one can in particular compute:
\begin{gather}\label{eq007}
\abs{\tilde{F}}^2 = \frac{8}{27} e^{-\phi} \left( f^2 - \frac{108}{5} m^2 \, e^{2\phi} \right) \\
\label{eqfm} f^2 \geq \frac{108}{5} m^2 \, e^{2\phi} \:.
\end{gather}
The further non-trivial constraint one gets from the Bianchi identities is $|\alpha|=constant$.
Note that the Bianchi identities are crucial to obtain a solution of all the equations of motion.
\

From these results we can obtain a characterisation of the $SU(3)$ structure of these backgrounds:

\begin{equation} \label{eq045}
\begin{split}
dJ &= \frac{2}{3} f e^{\phi/4} \re\Omega \\
d\Omega &= -\frac{4i}{9}f e^{\phi/4} J\wedge J -i \, e^{3\phi/4} J\wedge \tilde{F} \:,
\end{split}
\end{equation}
Thus, the nonvanishing torsion classes of $X_6$ are:
\begin{equation} \begin{split} \label{eq018}
\mathcal{W}_1^- &= -\frac{4i}{9}fe^{\phi/4} \\
\mathcal{W}_2^- &= -i e^{3\phi/4}\tilde{F} \\
\end{split} \end{equation}
A manifold with such an $SU(3)$ structure is a special case  of a so-called \emph{half-flat} manifold. (Compactifications on \emph{half-flat} manifolds are considered in \cite{Gurrieri:2002wz,Gurrieri:2002iw,Palti}).

\

From these results we can see that
the only Calabi-Yau solution (which has zero torsion) is the standard one with zero fluxes and zero cosmological constant.
The only other special class of solutions which can be considered have $\mathcal{W}_2^- = 0$ (because of \ref{eqfm}).
This requires $f^2 = \frac{108}{5}m^2 e^{2\phi}$. These manifolds are called \emph{nearly-K\"ahler}, and solutions of this kind were obtained in \cite{BehrCev}.

\section{IIA Supergravity with Orientifolds} \label{orientifold}

Our main result will be the ten dimensional description of the vacua discovered in \cite{deWolfe} (an example of such vacua is also given in \cite{Ihl}). Since these vacua must also have \hyph{O6}planes
we need to understand how the solutions of \cite{Lust} change in the presence of the $O6$ . The \hyph{O6}plane is not a genuine supergravity object, but rather something defined by the superstring compactification. Nevertheless, the supergravity action can be enriched with terms that describe the interactions of such an object with the low energy fields.

\

In IIA string theory, an orientifold 6-plane is obtained by modding out the theory by the discrete symmetry operator $\mathcal{O}$:
\begin{equation} \label{eq060}
\mathcal{O} \equiv \Omega_p (-1)^{F_L} \sigma^*
\end{equation}
where $\Omega_p$ is the world-sheet parity, $(-1)^{F_L}$ is the left-moving space-time fermion number, while $\sigma$ is an isometric involution of the original manifold.
The fixed point locus of $\sigma$ is the orientifold 6-plane.
In type IIA String Theory an \hyph{O6}plane is a BPS object, which preserves half of the supersymmetries: those such that $\epsilon_\pm = \mathcal{O} \, \epsilon_\mp$, where $\epsilon_\pm$ are the two Majorana-Weyl supersymmetry parameters \eqref{eq004}.

We are going to add an \hyph{O6}plane parallel to the $AdS_4$ factor, so three-dimensional in the internal manifold. Since the background preserves only four supercharges, in general an \hyph{O6}plane will break all of them. On the other hand, in order to get an \Nugual{1} four dimensional theory, we must take the $O6$ such that it preserves the same supercharges as the background. As in the case of a \hyph{D6}brane, this is achieved by wrapping the plane on a supersymmetric \hyph{3}cycle.

The operator $\mathcal{O}$ does not act on the four dimensional spinors $\theta_{\pm}$ while it exchanges $\eta_+$ and $\eta_-$.%
\footnote{Note that $\Omega_p (-1)^{F_L}$ acts trivially on the supersymmetry parameters, since they have the same parity properties of the metric.}
Thus
\begin{align}
\label{eq057}
J_{mn} = -i\eta_+^\dagger \gamma_{mn} \eta_+ &\stackrel{\sigma^*}{\longrightarrow} -i\eta_-^\dagger \gamma_{mn} \eta_- = -J_{mn} \\
\label{eq026}
\Omega_{mnp} = \eta_-^\dagger \gamma_{mnp} \eta_+ &\stackrel{\sigma^*}{\longrightarrow} \eta_+^\dagger \gamma_{mnp} \eta_-
        = -\Omega^*_{mnp}
\end{align}
Supersymmetry forces $\sigma$ to be antiholomorphic with respect to the almost complex structure $J$.

The fixed locus of the isometry $\sigma$ (if any) on the internal manifold is the supersymmetric \hyph{3}cycle $\Sigma$ the $O6$ wraps. In particular, we get for the pull-back to the plane:
\begin{equation}
 J|_\Sigma = 0 \qquad \qquad \re\Omega |_\Sigma = 0 \:,
\end{equation}
which implies
\begin{equation}
J \wedge \delta_3 = 0 \qquad \qquad \re\Omega \wedge \delta_3=0 \:.
\end{equation}
Moreover $\Omega$ is a calibration and $\Sigma$ is calibrated with respect to $-\im\Omega$. In fact one can compute
\begin{equation} \label{eq052}
\int_\Sigma \im\Omega = \im\Omega \wedge \delta_3 = - \frac{\delta^{(3)}(\Sigma)}{\sqrt{g_3^t}}\, \dvol = - \mbox{Vol}_\Sigma \:.
\end{equation}
These indeed show that $\Sigma$ is a supersymmetric \hyph{3}cycle (in fact special Lagrangian) \cite{CascUranga}.

One obtains the spatial parity of the other form fields by considering their worldsheet origin and imposing them to be invariant under the orientifold operator \eqref{eq060}: so, under $\sigma^*$, $F$ and $H$ are odd as $\delta_3$, $G$ is even. 

\

Now consider the modifications to the equations of motion (EOM) and the Bianchi identities (BI) given by the \hyph{O6}plane to Type IIA massive supergravity.
The bosonic action
is, at leading order in $\alpha'$:
\begin{equation} \label{eq002}
S_{O6} = 2 \mu_6 \int_{O6} d^7\xi e^{3\phi/4} \sqrt{-g_7} - 4 \mu_6 \int_{O6} C_7 \:,
\end{equation}
where the first piece comes from the Born-Infeld action, the second one from the Wess-Zumino's.%
\footnote{This action is directly derived from the one of a \hyph{D6}brane noticing that the orientifold projection forces $B$ to vanish on the plane, and \hyph{O}planes do not support gauge fields.}
Moreover $g_7$ is the pulled-back metric determinant on the plane, $\mu_6 = 2\kappa_{10}^2 \bar\mu_6= 2 \pi \sqrt{\alpha'}$, while $\bar\mu_p=(2\pi)^{-p}\alpha'^{-(p+1)/2}$ is the \hyph{Dp}brane charge and tension, and we have taken into account that the charge of an \hyph{Op}plane is $-2^{p-5}$ times that of a \hyph{Dp}brane.

These terms are only the first ones in an infinite expansion in $\alpha'$. Keeping just them and working with the leading supergravity action \eqref{eq003b} is consistent. In \Nugual{2} $10d$ supergravity theories, the first corrections coming from string theory are of order ${\alpha'}^3 R^4$, where $R^4$ stands for various contractions of four Rienmann tensors, to be compared to the leading term $R$.%
\footnote{For \Nugual{1} $10d$ theories the first corrections are of order $\alpha' R^2$.}
The orientifold leading action is instead of order $\sqrt{\alpha'}$. Classical solutions will be reliable only in regions where $\alpha' R \ll 1$.

The Born-Infeld term gives a contribution to the Einstein and dilaton equations, while the Wess-Zumino term represents an electric coupling to $C_7$.
The Born-Infeld term brings a localised contribution to the energy momentum tensor
\begin{equation}
 T_{MN}^{loc} \equiv -\frac{2}{\sqrt{-g}} \, \frac{\delta S_{O6}}{\delta g^{MN}}
 = 2\mu_6 \, e^{3\phi/4} \, \Pi_{MN} \, \frac{\delta^{(3)}(O6)}{\sqrt{g_3^t}} \:,
\end{equation}
where $\Pi_{MN}$ is the projected metric on the plane and $g_3^t=g_{10}/g_7$ is the determinant of the transverse metric. In case of a warped product metric as in \eqref{eq021} and for a submanifold wrapping the four-dimensional factor, $\Pi_{\mu\nu} = g_{\mu\nu}$.

The equations of motion are%
\footnote{Remember: ${F_p}^2= p! \abs{F_p}^2$. Moreover the equation of motion for $A$ is given by the differential of \eqref{eq051}.}
\begin{align}
\begin{split}
0 &= R_{MN} - \frac{1}{2}\partial_M\phi\partial_N\phi -\frac{1}{12}e^{\phi/2}G_M\cdot G_N
    +\frac{1}{128}e^{\phi/2}g_{MN} G^2  \\
 &\quad -\frac{1}{4}e^{-\phi}H_M\cdot H_N +\frac{1}{48}e^{-\phi}g_{MN} H^2
    -\frac{1}{2}e^{3\phi/2}F_M\cdot F_N+\frac{1}{32}e^{3\phi/2}g_{MN} F^2
       \\
 &\quad -\frac{1}{4}m^2e^{5\phi/2}g_{MN} -\mu_6 e^{3\phi/4} \Pi_{MN} \frac{\delta^{(3)}(O6)}{\sqrt{g_3^t}}
    +\frac{7}{8}\mu_6 e^{3\phi/4} g_{MN} \frac{\delta^{(3)}(O6)}{\sqrt{g_3^t}}
\end{split} \\
\label{eq056}
\begin{split}
0 &= \nabla^2\phi -\frac{1}{96}e^{\phi/2} G^2
        +\frac{1}{12}e^{-\phi}H^2
 -\frac{3}{8}e^{3\phi/2} F^2  -5m^2e^{5\phi/2} \\
 &\quad  +\frac{3}{2}\mu_6 e^{3\phi/4} \frac{\delta^{(3)}(O6)}{\sqrt{g_3^t}}
\end{split} \\
\label{eq051}
0 &= d(e^\phi \ast H) - \frac{1}{2} G\wedge G +  e^{\phi/2} F\wedge \ast G + 2m e^{3\phi/2} \ast F \\
\label{eq030}
0&= d(e^{\phi/2} \ast G) - H\wedge G \:.
\end{align}
Here $X_M\cdot X_N$ means contraction on all but the first index.
Notice that the only equations that get modified with respect to \cite{Lust}, due to the presence of an orientifold plane, are the Einstein and dilaton equations.

The Wess-Zumino term in \eqref{eq002} describes the coupling of the plane to $C_7$, which is the gauge potential dual to $A$, and so the $O6$ is a magnetic source for $A$. This term does not modify the equations of motion, but only the Bianchi identity. The way this modification can be evaluated is taking the dual description in terms of $F_8$, so that the BI is obtained by varying with respect to $C_7$. We obtain
\begin{equation} \label{eq010}
dF = 2m H - 2 \mu_6\, \delta_3 \qquad \qquad dH = 0 \:.
\end{equation}
The other BI is $dG =F\wedge H$ is satisfied.%
\footnote{Looking at the complete Wess-Zumino term for a $D$6-brane, one could have suspected a localized modification to the BI for $G$ like $\delta_3\wedge F$. But the orientifold projection forces the pull-back of $F$ on the plane to vanish. This would not necessarily be true for \hyph{D}6-branes.}

In the derivation it has been convenient to express integrals on the plane as integrals on the whole space, through the \hyph{3}form $\delta_3$, transverse to the plane and localized on it:
\begin{equation} \label{eq050}
\int_{O6} C_7 = \int C_7 \wedge \delta_3 \:.
\end{equation}
In local coordinates $y_M$, where the \hyph{O6}plane is located ad $y^7=\ldots=y^9=0$, we have
$\delta_3 = \delta^{(3)} (y^7,y^8,y^9) \: dy^7 \wedge dy^8 \wedge dy^9$
expressed through a usual delta function.
Notice the closure
\begin{equation}
d\delta_3 = 0 \:,
\end{equation}
which means nothing more than charge conservation. A precise treatment of distributional forms would be to consider the embedding of a seven dimensional manifold $M_7$ into the target space $f:M_7 \to Z$, so that $\int_{M_7} f^* C_7$ is a nondegenerate linear map from \hyph{7}forms to real numbers. The Poincar\'e dual to $f(M_7)$ is now, by definition, an object $\delta_3$ which realizes \eqref{eq050} as a linear map on \hyph{7}forms. It turns out that the differential $d\delta_3$ is defined by $\int C_6 \wedge d\delta_3 = - \int_{\partial M_7} f^* C_6$ on \hyph{6}forms. In our case the \hyph{O6}plane has no boundary, hence closure.

\

Summarizing, the introduction of the \hyph{O6}plane does not modify the SUSY variations in \eqref{eq009}; it changes the Bianchi identity for the \hyph{2}form field-strength and
induces some additional terms in the Einstein and dilaton equations of motion.

In order to find the new solution, we follow the same procedure as in \cite{Lust}, i.e. we solve the SUSY equations
$\delta\psi_M=0$ and $\delta\lambda=0$, and then we impose BI's and EOM's for form fields.
In fact, one can show that the Einstein and dilaton equations are automatically satisfied (a part from the minor requirement on the Einstein equation $E_{0M}=0$ for $M\neq 0$, which is granted with the ansatz \eqref{eq021}). We will partly verify it in the appendix \ref{check}.

The system of relations \eqref{eq006}  solve also the form field equations \eqref{eq051}, \eqref{eq030} and the BI for $G$. So we are left with only the modified BI for $F$ \eqref{eq010}.Substituting the solution (\ref{eq006}) into the modified BI and using the expression
\eqref{eq045} for $dJ$, one gets
\begin{equation} \label{eq027}
 d\tilde{F}= - \frac{2}{27} e^{-\phi/4}\left( f^2 - \frac{108}{5} m^2 e^{2\phi} \right)
            \re\Omega - 2 \mu_6 \, \delta_3 \:.
\end{equation}
From this we can compute $\abs{\tilde F}^2$. Start from $0 = d(\Omega\wedge \tilde F)$, use again \eqref{eq007} and \eqref{eq052} to get
\begin{equation} \label{eq028}
 \abs{\tilde{F}}^2 = \frac{8}{27} e^{-\phi}\left( f^2 - \frac{108}{5} m^2 e^{2\phi} \right)
        + 2 \mu_6 e^{-3\phi/4} \frac{\delta^3(\Sigma)}{\sqrt{g_3^t}} \:.
\end{equation}
The first term is constant on $X_6$, while the second one has support on the cycle
$\Sigma$. $\abs{\tilde{F}}^2$ is positive definite, so we find two conditions:
\begin{equation} \label{eq019}
f^2 \geq \frac{108}{5}m^2 e^{2\phi} \qquad \mbox{and} \qquad \mu_6 \geq 0 \:.
\end{equation}
Note that the latter is perfectly expected: changing the sign of the charge of the \hyph{O6}plane gives an anti-\hyph{O6}plane, which however preserves orthogonal supersymmetries incompatible with the background.
The discussion of the possibility of getting a Calabi-Yau geometry is parallel to section \ref{LustSol}. One would have to put $f$ and $\tilde F$ to zero, but this would also imply $m$ vanishing. The massless limit has to be taken with care, and one finds Calabi-Yau without flux. Moreover, as long as the localized contribution is present, there will always be a singular behaviour on it, captured by \eqref{eq045}.

\subsection{A Smeared Solution} \label{smeared}

To find exact solutions in presence of localized objects is not easy, mainly because, as we saw, {\it in no case with non vanishing mass parameter does the geometry reduce to Calabi-Yau}. Nevertheless, as a first step, we can consider a long-wavelength approximation in which this situation is realized. In a Calabi-Yau metric the torsion classes vanish:
\begin{equation}
    f=0 \qquad \tilde F=0 \qquad F=0 \qquad m^2 > 0 \:.
\end{equation}
In the long-wavelength approximation the charge of the orientifold plane, localized on $\Sigma$, is substituted with a smeared distribution (obviously keeping the total charge the same).
Thus the \hyph{3}form describing the new charge distribution must be in the same cohomology class as $\delta_3$. Integrating the Bianchi identity \eqref{eq010} on \hyph{3}cycles gives the tadpole cancellation conditions, and the solution \eqref{eq006} immediately suggests a natural choice for the smooth 3-form:
\begin{equation} \label{eq053}
    \mu_6 \, \delta_3 \longrightarrow \frac{4 m^2}{5} \, e^{7\phi/4} \re\Omega \:.
\end{equation}
Direct inspection of \eqref{eq027} shows that in fact we can consistently put $f$ and $\tilde F$ to zero.

On the other hand, we can find which is the Poincar\'e dual to the cycle $\Sigma$ on a Calabi-Yau manifold, without invoking physical constraints like BI's. Let us consider a symplectic basis for homology $\{\Sigma_A,\Gamma^B\}$ with $A,B=0,\dots,h^{2,1}$, intersection numbers $\Sigma_A \cap \Gamma^B = \delta_A^B$ and such that $\Sigma_0 \equiv \Sigma$ the orientifold locus. We will call the cycle $\Gamma^0$ just $\Gamma$. Then we construct the dual basis of integral cohomology classes $\{\alpha_K, \beta^L\}$ defined by
\begin{equation}
    \int_{\Sigma_A}\alpha_K = \delta_K^A \qquad \int_{\Gamma_B}\beta^L = \delta_L^B
\end{equation}
while the other combinations vanishing. It turns out that%
\footnote{From here one derives a formula for the volume of this sLag \hyph{3}cycle: $\mbox{Vol}_\Sigma = \sqrt{4\vol}$.}
\begin{equation} \label{eq062}
\{\Sigma,\Gamma\} \leftrightarrow \{\alpha_0 = - \frac{\im\Omega}{\sqrt{4\vol}} , \beta^0 = \frac{\re\Omega}{\sqrt{4\vol}} \}
\end{equation}
with $\alpha_0,\beta^0$ respectively even and odd under $\sigma^*$, and the other dual forms extracted from $H^{2,1} \oplus H^{1,2}$. This normalization comes down from the choice \eqref{eq025}. Since the only nonvanishing period of $\delta_3$ is $\int_{\Gamma} \delta_3 = 1$ (because $\Sigma$ and $\Gamma$ intersect just once), we find that the harmonic representative of the cohomology class of $\delta_3$ is
\begin{equation} \label{eq066}
\delta_3 \longrightarrow \beta^0 = \frac{\re\Omega}{\sqrt{4\vol}} \:.
\end{equation}

Comparing with \eqref{eq053} we get the value of the dilaton:
\begin{equation} \label{eq054}
\frac{4 m^2}{5} \, e^{7\phi/4} = \frac{\mu_6}{\sqrt{4 \vol}} \:.
\end{equation}
This fixes also the value of the four-dimensional cosmological constant. Summarizing, the solution is completely described by the internal Calabi-Yau manifold defined by \hyph{SU(3)}invariant forms $J$ and $\Omega$, with an anti-holomorphic isometrical involution $\sigma$: the background fields $G$ and $H$ are determined by \eqref{eq006} with $f=0$, $F=0$; the dilaton is given by \eqref{eq054} where in turn the volume is set by $J$. Further constraints come from the integral quantization of fluxes, and this mechanism provides the stabilization of geometrical moduli in the geometry. This will be analyzed in the next section.

It would be of interest to establish in even more detail how the smeared and localised exact solutions
are related.

\subsection{Tadpole Cancellation and Topology Change}

In the exact localised solution, the fact that $\re \Omega$ is exact implies that
$H$ must be exact.
The most important consequence is that the modified BI implies that $mH - \sum_i \mu_6 \delta^{(i)}_3$ must vanish in cohomology; here $i$ runs over all the localized sources. Therefore from the tadpole cancellation conditions one gets that the possible configurations of localized charges are constrained: charge cancellation must work among localised charges only. Specifically, it must be that:
\begin{equation} \label{eq049}
\int \sum_i \delta^{(i)}_3 = 0
\end{equation}
on all closed \hyph{3}cycles. 
This is different from the smeared CY solution (in which $f=0$), where a non-trivial closed $H$ was allowed by the supersymmetry equations and could be used to cancel the $O6$ charge.

In the case of a single source we see that $\delta_3$ is exact. Since $\delta_3$ is the Poincare dual
of the homology class of the \hyph{O6}plane, we learn that the 3-cycle that the \hyph{O6}plane wraps is contractible.
This is in stark contrast to the smeared Calabi-Yau case in which the \hyph{O6}plane is necessarily non-trivial
in homology. Therefore, we learn that the transition from the Calabi-Yau approximation to the exact solution necessarily involves a topology change.

\section{Moduli Stabilization} \label{stabilization}

In this section we will describe from the point of view of ten dimensional supergravity, how the introduction of the fluxes
stabilise the moduli which are present in the zero flux, Calabi-Yau limit. After a brief general discussion, we will first discuss the
moduli vevs in the examples studied in \cite{deWolfe} and then go on to discuss the general case.

\

We begin with the axions. A background value for the field strength of a gauge form potential can be separated in two pieces:
\begin{equation} \label{eq031}
H = H^f + dB \:.
\end{equation}
The former, cohomologically nontrivial, when integrated on cycles gives the integer amounts of flux, whilst the second term is globally exact. 
$H^f$ must be closed (so that the flux depends only on co-homology), and we can {\it choose} an harmonic representative of the integral cohomology class. Note however that this separation is arbitrary. From the exact solution the total field strength $H$ is harmonic so that $dB=0$.
We can then use the gauge freedom $B \to B + d\lambda$ to choose $B$ harmonic. The internal harmonic compopnents of $B$ are four dimensional
axions. This shows that all the other Kaluza-Klein modes have a zero vacuum expectation value and are hence massive.

In the same way, we split the other field-strengths:
\begin{align}
\label{eq036}
F &= F^f +  dA + 2 m\,B \\
\label{eq032}
G &= G^f + f d\mbox{Vol}_4 + dC + B\wedge dA + mB^2 \:.
\end{align}
Arguing as before, ${F}^f$ is the integrally quantized flux of the gauge potential $A$ while $G^f$ is the flux of $C$; all of them can be taken harmonic exploiting the gauge redundancy. Note that being $A$ harmonic, it is actually vanishing on our Calabi-Yau solution because of the vanishing of $H^1(CY, {\mathbb R})$.

So one simply expands the fluxes (quantized), the gauge potentials and the \hyph{SU(3)}structure forms defining the metric. The right basis is dictated by the exact solution, and by the constraints imposed by the orientifold projection. In the special example at hand, everything is harmonic. On the other hand, we can only study the vacuum and can not go off-shell, so can not see any superpotential.

In order to discuss the stabilization of axions coming from $C$, we need to consider the BI for $\tilde F_6 \equiv e^{\phi/2} \ast_E G$, or equivalently the EOM \eqref{eq030}. Splitting the field strength according to \eqref{eq031} and \eqref{eq032} and recalling that $A=0$ one can recast it in the form of an exact differential:
\begin{equation}
d\big( e^{\phi/2} \ast G + H\wedge C - B \wedge G^f - \frac{1}{3} m\, B^3 \big) = 0 \:.
\end{equation}
When $f \neq 0$, $C$ must contain also a four-dimensional piece $C_M$ such that $dC_M = f d\mbox{Vol}_4$. Being a BI, the term in parenthesis is recognized as the closed component of $\tilde F_6$, which can be further split into flux and an exact piece:
\begin{equation} \label{eq037}
F_6^f + dC_5 = e^{\phi/2}\ast G + H \wedge C - B \wedge G^f - \frac{1}{3}m\, B^3 \:.
\end{equation}

\subsection{Example: the $T^6 / (\mathbb{Z}_3)^2$ Orientifold}

The smeared solution in the long-wavelength approximation can be exploited to compare results with another widely used approximation: what is called Calabi-Yau with fluxes. In the latter, one keeps the contribution of fluxes small compared to the curvature of the compactification manifold. Note that fluxes can not be taken arbitrarily small; Dirac quantization condition puts a lower bound $F_p \sim (\alpha')^{\frac{p-1}{2}}$ to the amount for a \hyph{p}field-strength. So one requires the contribution of fluxes to the action to be small compared to the Einstein term $R$, which is of order $L^{-2}$ with respect to the characteristic length of the manifold. This gives $(\alpha'/L^2)^{p-1} \ll 1$. In other words, we must be in the limit of large compactification manifold with respect to the string length, which anyway is the regime of applicability of supergravity. Under these conditions, one can neglect the backreaction of fluxes on geometry, and work with the Calabi-Yau metric. Of course one has to be careful to remember that in the action there are factors of the dilaton, and both the dilaton and the volume are (possibly) determined by fluxes themselves, so it is not always possible to keep the fluxes to their minimal amount while increasing the volume. On the other hand, the
smeared solution is valid for large flux.

A simple example studied in detail by \cite{deWolfe} is the $T^6 / {\mathbb{Z}_3}^2$ orientifold and will be
useful as a concrete model.
The model is constructed by compactifying Type IIA supergravity on a \hyph{6}manifold which is (the singular limit of) a Calabi-Yau: a torus $T^6$ firstly orbifolded by ${\mathbb{Z}_3}^2$ and then orientifolded. It has Hodge numbers $h^{2,1}=0$ and $h^{1,1}=12$, where 9 of the 12 K\"ahler moduli arise from the blow-up modes of 9 $\mathbb{Z}_3$ singularities. There are no complex structure moduli.
The \hyph{O6}plane wraps a special Lagrangian \hyph{3}cycle and is compatible with the closed \hyph{SU(3)}structure of the CY. The resulting theory has 4 preserved supercharges. The number of moduli from the form fields are: 3 from the NS-NS \hyph{2}form potential $B$ (odd under $\sigma$), no one from the R-R \hyph{1}form potential $A$ and 1 from the R-R \hyph{3}form potential $C$ (even).
Fluxes are switched on as described above.

In \cite{deWolfe} the stabilization of the moduli, due to the fluxes, is analysed by a computation of the four dimensional effective moduli potential. We are going to apply to this model the machinery previously developed, in the long-wavelength approximation.

\

Let us introduce an integer basis of harmonic forms for the even cohomology groups. The \hyph{2}forms (odd under $\sigma$) $w_i$:
\begin{equation}
w_i \propto \frac{i}{2}dz_i\wedge d\bar z_i \qquad \qquad  \int w_1 \wedge w_2 \wedge w_3 = 1 \:.
\end{equation} 
The \hyph{4}forms (even under $\sigma$)
\begin{equation}
\tilde w^i = w_j \wedge w_k \qquad \Rightarrow \qquad \int w_a \wedge \tilde w^b = \delta_a^b
\end{equation} 
where $j$ and $k$ are the two values of $1,2,3$ besides $i$.

Start with the decomposition of $F$ \eqref{eq036}. Expand the fields on harmonic forms (of correct parity)
\begin{equation} \label{eq034}
{F}^f = f^i \, w_i \qquad \qquad B = b^i \, w_i \:,
\end{equation} 
where $f^i$ are quantized in units of $\mu_6$. Imposing the smeared solution $F=0$, we get
\begin{equation}
b^i = - \frac{f^i}{2m} \:.
\end{equation} 
The ``moduli''%
\footnote{We call them moduli because they are so in the Calabi-Yau compactification without fluxes, but here the exact solution fixes completely $B$, and so there are no moduli at all.}
$b^i$ corresponding to four dimensional axions are fixed by the fluxes $f^i$. We can take for simplicity ${F}^f=0$, as in \cite{deWolfe}, then $B=0$ and the axions are fixed to $b^i=0$. The general case is dealed for in the next section.

Then expand the \hyph{4}form flux $G$ and the \hyph{SU(3)}structure foundamental form
\begin{align}
\label{eq035}
G^f &= \sum_i e_i \, \tilde w^i \\
\label{eq033}
J &= e^{-\phi/2} \sum_i v^i \, w_i \qquad v^i > 0 \:,
\end{align} 
where $e_i$ are quantized in units of $\mu_4$, and we put a power of the dilaton for later convenience. Note in particular
\begin{equation} \label{eq016}
v^1 v^2 v^3 = e^{3\phi/2} \, \vol = \vol^\textit{String frame} \:.
\end{equation}
Substituting into the decomposition of $G$ \eqref{eq032} and in the solution \eqref{eq006} with $f=0$ and $b^i=0$, we get
\begin{equation}
\frac{6m}{5} \, v^j v^k = e_i \:.
\end{equation}
We find a series of relations on the possible fluxes that characterize a supersymmetric vacuum: $\sgn (m\,e_1e_2e_3) = \sgn (m\, e_i) = +$ and the sign of $e_i$ is independent on $i$. These are in agreement with \cite{deWolfe}. Moreover we can invert to
\begin{equation} \label{eq015}
v^i = \frac{1}{\abs{e_i}} \sqrt{\frac{5}{6} \, \frac{e_1e_2e_3}{m}} \:.
\end{equation} 
So the K\"ahler moduli are fixed. In the more general case $b^i \neq 0$ they are still fixed, a part from changing the range of fluxes for which the supergravity approximation is reliable.

The stabilization of the dilaton comes from the decomposition of $H$ \eqref{eq031}. Expand $H$ in a basis of harmonic forms for the third cohomology group, odd under the spatial orientifold operation $\sigma^*$. In the present example there is only $\re\Omega$. Note that this is consistent with the solution \eqref{eq006}. So let us put
\begin{equation}
H = H^f = p \frac{1}{\sqrt{4 \vol}} \re\Omega \:.
\end{equation} 
The normalization comes from $\int_\Gamma \beta^0 = 1$ (see also \eqref{eq062}), so $p$ is integrally quantized in units of $\mu_5$.
Integrating the BI for $H$ on the cycle $\Gamma$ we get the only nontrivial tadpole cancellation condition
\begin{equation}
\int_\Gamma m\,H = m\,p= \mu_6
\end{equation} 
whose only two solutions are%
\footnote{Note, in quantizing $m$, that it is not canonically normalized in the action \eqref{eq003b}; then it is quantized in units of $\mu_8/2$.}
$(m,p) = \pm (\mu_8/2, 2 \mu_5)$ and $\pm(\mu_8, \mu_5)$. Comparing with the solution, the dilaton gets stabilized to
\begin{equation}
e^\phi = \frac{3}{4} \mu_6 \left( \frac{5}{6} \, \frac{1}{m^5\,e_1e_2e_3} \right)^{1/4} \:.
\end{equation} 

The last issue is the stabilisation of possible axions coming from the \hyph{3}form potential $C$. Being it odd under $\sigma^*$ and harmonic, there is only one axion:
\begin{equation}
C = - \xi \frac{\im \Omega}{\sqrt{4 \vol}} \:.
\end{equation} 
This must be substituted into the decomposition of the field-strength $\tilde F_6$ dual to $G$ \eqref{eq037}, with quantized flux $\int F_6^f = e_0$. We get:
\begin{equation}
-p \, \xi = e_0 
\end{equation} 

The result is that, in this simple model, all the K\"ahler moduli, the dilaton and the only axion are geometrically stabilized, whilst there are no complex structure moduli. {\it All the results found in this section are in precise agreement with those found in \cite{deWolfe}}. Really one should discuss the moduli associated to the 9 resolved singularities as well, which are one K\"ahler modulus each. One would find that the singularities are blown up to a finite volume. In the next section will discuss how this example generalizes to any Calabi-Yau, of which the orbifold is just a singular limit.

\

We can determine also the four-dimensional cosmological constant, that is the vacuum energy in $AdS_4$. The exact solution \eqref{eq006} gives the scalar curvature $\hat R=-24\abs{W}^2$ of the $AdS_4$ factor in ten dimensional Einstein metric (note that the constant $\Delta$ cancells out). Then we must express it in four dimensional Einstein frame, through
\begin{equation}
R^{4DE} = M_P^2  \kappa_{10}^{2} \frac{1}{\vol} \hat{R} = -\frac{24}{25} M_P^2 \kappa_{10}^{2} m^2 \frac{e^{5\phi/2}}{\vol} \:.
\end{equation} 
Eventually, choosing conventions for the Einstein equation $R_{\mu\nu} -\frac{1}{2} g_{\mu\nu} R = -\frac{1}{2} g_{\mu\nu} \Lambda$:
\begin{equation}
\Lambda = - (2\pi)^{11} \left( \frac{3}{4} \right)^4  \left( \frac{6}{5} \frac{\alpha'^4}{m\,e_1e_2e_3}  \right)^{3/2} \, M_P^2 \:.
\end{equation}

\subsection{General Calabi-Yau with Fluxes}

The generalization of this example to any Calabi-Yau model with an orientifold projection is straightforward. 
We will continue to adopt the long-wavelength approximation as done in the previous section.
First of all the antiholomorphic involutive isometry $\sigma$ divides the cohomology groups of the internal manifold into even and odd components. In particular, $H^{1,1}= H^{1,1}_+ \oplus H^{1,1}_-$ with dimensions $h^{1,1} = h^{1,1}_+ + h^{1,1}_-$. Let $\{ w_i \}$ be an integer basis for $H^{1,1}_-$, with intersection numbers
\begin{equation}
\kappa_{abc} = \int w_a \wedge w_b \wedge w_c \:,
\end{equation} 
and $\{ \tilde w^i \}$ the Poincar\'e dual basis for $H^{2,2}_+$ (since $J^3$ is odd):
\begin{equation}
\int w_i \wedge \tilde w^j = \delta_i^j \:.
\end{equation}
The third cohomology group $H^3 = H^3_+ \oplus H^3_-$ is halved in two spaces of real dimension $h^{2,1}+1$. We consider the same integer symplectic real basis for $H^3$ as in section \ref{smeared}: $\{\alpha_K, \beta^L\}$ with $k,l:0,\ldots,h^{2,1}$. It satisfies $\int \alpha_K \wedge \beta^L = \delta_K^L$;
moreover $\alpha_K$ are even while $\beta^L$ are odd. Let the dual basis of integer cycles be $\{ \Sigma_A, \Gamma^B \}$ so that $\Sigma_A \cap \Gamma^B = \delta^B_
A$.

Then we expand the various fields and forms on these basis, according to their behaviour under the orientifold operation $\mathcal{O}$. The K\"ahler form $J$, the field $B$ and the flux $F^f$ are odd and follow \eqref{eq033}, \eqref{eq034}.%
\footnote{A possible axion coming from $B$ lying on the four dimensional space is forbidden by the orientifold projection.}
In particular
\begin{equation} \label{eq038}
\vol = \frac{1}{6} e^{-3\phi/2} \, v^a v^b v^c \, \kappa_{abc} \:.
\end{equation} 
The flux $G^f$ is even and follows \eqref{eq035}. The treatment of the holomorphic \hyph{3}form needs a little bit more of care. On a Calabi-Yau it can be expanded on the full $H^3$:
\begin{equation}
\Omega = g^K \alpha_K + Z_L \beta^L \:.
\end{equation} 
We can take $Z_L$ as projective coordinates on the complex structure moduli space of the Calabi-Yau, while $g^K$ as functions of $Z_L$ on this space. Nonetheless, we choose the particular normalization $\Omega\wedge \bar\Omega = -8i\dvol$, and this fixes the overall factor. Then the orientifold projection requires $\re\Omega$ and $\im\Omega$ to be respectively odd and even under $\sigma$; this translates to
\begin{equation}
\im Z_L = \re g^K = 0 \:.
\end{equation}
Notice that while the first set of relations really cuts out half of the moduli space, the second set is automatically garanteed on a CY manifold which admits the antiholomorphic isometry $\sigma$.
The flux $H^f$ is odd and the gauge potential $C$ is even, so
\begin{equation}
H=H^f= p_L \beta^L \qquad \qquad C= \xi^K \alpha_K \:.
\end{equation}

The stabilization proceeds on the same track as before. We substitute the expantions given above in the equations determining the solution. From \eqref{eq036} and \eqref{eq032} we get
\begin{gather} \label{eq065}
b^i = - \frac{f^i}{2m} \\
\frac{3m}{5} v^i v^j \, \kappa_{ija} = e_a + m \, b^i b^j \, \kappa_{ija} \:.
\end{gather} 
The axions $b^i$ are all fixed, as well as the K\"ahler moduli $v^i$. For these last ones we have as many quadratic equations as unknowns (provided thare is no $a$ such that $\kappa_{aij}$ is always zero), and, as pointed out in \cite{deWolfe}, one has only to check that the solution lies in the supergravity regime (among the others, one asks for large positive volumes $v^i$). Integrating the BI for $H$ on the cycles $\Gamma_L$ yields
\begin{equation}
m\,p_L = \mu_6 \frac{\re Z_L}{\sqrt{4\vol}} \:.
\end{equation}
This fixes all the remaining complex structure moduli%
\footnote{The equations are not invariant under scaling (what one would have expected for the projective coordinates), but this relies on the fact that a normalization for $\Omega$ is involved in \eqref{eq066}.}%
. Then subsituting in the solution \eqref{eq006} we find the dilaton
\begin{equation} \label{eq043}
e^\phi = \frac{5}{8} \frac{\mu_6}{m^2} \sqrt{ \frac{6}{v_a v_b v_c \, \kappa_{abc}} } \:.
\end{equation} 
Eventually, by direct application of \eqref{eq037} follows
\begin{equation}
-p_L \, \xi_L = e_0 + b_i e_i + \frac{1}{3} m \, b_a b_b b_c \, \kappa_{abc} \:.
\end{equation} 
Note that only this particular combination of the axions can be fixed, while for the other ones nonperturbative effects and $\alpha'$ corrections must be invoked. Anyway, the stabilization of axions is a minor problem, because their configuration space is periodic and compact, so any contribution which generate a nonconstant potential fixes them at a finite value.

\

As noted in \cite{deWolfe}, there is a gauge redundancy in the solutions described above, i.e. solutions which are transformed into each other by the gauge transformations \eqref{eq064} and following, are equivalent. In the four-dimensional low energy theory those translate in Peccei-Quinn symmetries that shift the axions:
\begin{equation}
b^i \to b^i + 1 \qquad \mbox{or} \qquad \xi^K \to \xi^K + 1 \:.
\end{equation}
These are accompained by translations of the fluxes, and the correct transformation rules are obtained by \eqref{eq036}, \eqref{eq032} ,\eqref{eq037} by noticing that $F$, $G$ and $F_6$ are gauge-invariant.
The point is that one can always reduce to the case of $b^i$ and $\xi^K$ of order unity, and the large volume limit (the one reliable in supergravity) is controlled just by the fluxes $e_i$. This simplifies considerably the equations in the limit.

\

As in the particular case studied in the previous section, we have found the same results as \cite{deWolfe}: all the geometric moduli and the axions coming from $B$ are fixed, whilst only one combination of the $C$ axions is fixed.

\begin{center} \textbf{Acknowledgments} \end{center}

We would like to thank Matteo Bertolini, Stefano Cremonesi, Mariana Gra\~na, Giuseppe Milanesi, Marco Serone for useful discussions, and Thomas House, Eran Palti for some correspondence.

\appendix

\section{Ten Dimensional Action and Supesymmetry Variations} \label{action}

The bosonic action of the Type IIA massive supergravity \cite{Romans} with mass parameter%
\footnote{In string theory, this parameter is really a flux $F_0$, in fact quantized.}
$m$ is given, in Einstein frame, by%
\footnote{In order not to clutter formulas, we omit a factor $1/2\kappa_{10}^2=(2\pi)^{-7}\alpha'^{-4}$ in front of the lagrangian. But to discuss the supergravity limit and the various orders in $\alpha'$, this term has to be taken into account, and not just put to one.}
\begin{multline} \label{eq003b}
\mathcal{L} = \int \big\{ R \ast 1 - \frac{1}{2} d\phi \wedge \ast d\phi - \frac{1}{2} e^{\phi/2} G \wedge \ast G - \frac{1}{2} e^{-\phi} H \wedge \ast H \\
- \frac{1}{2} e^{3\phi/2} F \wedge \ast F - 2m^2 e^{5\phi/2} \ast 1 + \frac{1}{2}d C^2 \wedge B + \frac{1}{2} d C\wedge dA\wedge B^2 \\
+ \frac{1}{6}dA^2\wedge B^3 + \frac{m}{3} d C\wedge B^3 + \frac{m}{4}dA\wedge B^4 + \frac{m^2}{10}B^5  \big\} \:,
\end{multline}
where the invariant field strength with their BI's are:
\begin{equation} \label{eq023} \begin{split}
F &= dA + 2mB   \\
H &= dB               \\
G &= d C + B \wedge dA + m B^2
\end{split} \qquad \qquad
\begin{split}
dF &= 2m H \\
dH &= 0 \\
dG &= F \wedge H \:.
\end{split}
\end{equation} 
The gauge transformations which leave the action invariant are:
\begin{equation} \label{eq064}
\delta A = m \Lambda_1 \qquad
\delta B = -\frac{1}{2} d\Lambda_1 \qquad
\delta C =  \frac{1}{2} A\wedge d\Lambda_1 + \frac{1}{4} m \Lambda_1\wedge d\Lambda_1\:,
\end{equation}
as well as $\delta A = d \Lambda_0$ and $\delta C=d\Lambda_2$.

For a canonically normalized field-strength, the Dirac quantization condition states
\begin{equation}
\int_{\Sigma_p} F_p = \mu_{8-p} n_p = (4\pi^2\alpha')^\frac{p-1}{2} n_p \qquad \qquad n_p \in \mathbb{Z} \:,
\end{equation} 
with $\mu_p = 2 {\kappa_{10}}^2 \bar \mu_p = (4\pi^2\alpha')^{(7-p)/2}$, and  $\bar \mu_p = (2\pi)^{-p} \alpha'^{-(p+1)/2}$ is the \hyph{Dp}brane charge and tension.

The condition for a background to be supersymmetric, is that it satisfies the equations
\begin{equation}
\delta \Psi_M=0 \qquad \mbox{and} \qquad \delta\lambda=0
\end{equation}
where
\begin{align} \label{eq009}
\begin{split}
\delta \Psi_M &= \bigg[ \nabla_M - \frac{m \, e^{5\phi/4}}{16} \Gamma_M -
		\frac{e^{3\phi/4}}{64} F_{NP} ({\Gamma_M}^{NP} - 14{\delta_M}^N \Gamma^P) \Gamma_{11} \\
	&\quad +\frac{e^{-\phi/2}}{96} H_{NPQ} ({\Gamma_M}^{NPQ} - 9{\delta_M}^N \Gamma^{PQ}) \Gamma_{11} \\ 
	&\quad + \frac{e^{\phi/4}}{256} G_{NPQR} ({\Gamma_M}^{NPQR}
		- \frac{20}{3}{\delta_M}^N \Gamma^{PQR}) \bigg] \epsilon
\end{split} \\
\begin{split}
\delta\lambda &= \bigg[ -\frac{1}{2} \Gamma^M \nabla_M \phi - \frac{5m \, e^{5\phi/4}}{4} + 
                \frac{3 \, e^{3\phi/4}}{16} F_{MN} \Gamma^{MN} \Gamma_{11} \\ 
	&\quad +\frac{e^{-\phi/2}}{24} H_{MNP} \Gamma^{MNP} \Gamma_{11}  
                -\frac{e^{\phi/4}}{192} G_{MNPQ} \Gamma^{MNPQ} \bigg] \epsilon
\end{split}
\end{align}
In order to solve this, one substitutes the ansatz for $\epsilon$ \eqref{eq004}, for the metric and for the forms and contracts the resulting six dimensional equations with $\eta^\dagger_\pm \gamma^{(n)}$. In this way, one obtains separate equations for every $SU(3)$ representation in the decomposition of forms \cite{Lust}: one can decompose the tensors $F$, $H$ and $G$ in terms of irreducible $SU(3)$ representations.
For example, for $F$ one gets:
\begin{equation}
F_{mn}=\frac{1}{16}{\Omega^*_{mn}}^s F_s^{(1,0)}+\frac{1}{16}{\Omega_{mn}}^s F_s^{(0,1)}+
	(\tilde{F}_{mn}+\frac{1}{6}J_{mn}F^{(0)}) \:,
\end{equation}
where the different pieces can be extracted through
\begin{equation}
F^{(0)}=F_{mn}J^{mn} \sim {\bf 1} \qquad \qquad F_m^{(1,0)}={\Omega_{m}}^{np}F_{np} \sim {\bf 3}
\end{equation}
and $\tilde F\sim \mathbf{8}$ is such that
\begin{equation}
\tilde F_{mn} J^{mn} = \tilde F_{mn} {\Omega^{mn}}_p = \tilde F_{mn} {(\Omega^*)^{mn}}_p = 0 \:.
\end{equation} 
By different contractions one has a set of equations, and then recasting togheter the various pieces one gets (\ref{eq006}) (in case $|\alpha|=|\beta|$).

\section{Check of the Equations of Motion} \label{check}

In section \ref{orientifold} we sketched an argument to find that if the solution to the supersymmetry equations satisfies also the BI and the equations of motions for the forms, then it satisfies the Einstein and the dilaton equations as well. 
Here we check that it is true for the dilaton and the 4-dimensional components of the Einstein equation.

The dilaton eom \eqref{eq056} is the same as in \cite{Lust}, but with the addition of the $O6$ term. 
Moreover, the fields take the same values on the solution as in 
\cite{Lust}, except for $F$. The value of $F^2$ is the \cite{Lust} one plus 
\begin{equation} \label{eq011}
\delta F^2=\frac{1}{4}\mu_6 \frac{\sqrt{-g_3}}{\sqrt{-g_6}} \delta^3(\Sigma)e^{-3\phi/4} \:.
\end{equation}
So if the \cite{Lust} EOM are satisfied, all the terms in \eqref{eq056} 
sum up to zero, except for
\begin{equation}\label{eq012}
-\frac{3}{8} e^{3\phi/2} \delta \abs{F}^2 + 
	\frac{3}{2}\mu_6 \frac{\sqrt{-g_3}}{\sqrt{-g_6}} \delta^3(\Sigma)e^{3\phi/4} \:.
\end{equation}
By substituting \eqref{eq011} into \eqref{eq012} one gets exactly zero and the dilaton 
EOM turns out to be correct.

Consider, now, the Einstein EOM in the $\mu,\nu=0,...,3$ directions. The piece of the equation 
which is not automatically zero if the \cite{Lust} EOM are satisfied is:
\begin{equation}
\frac{1}{32}e^{3\phi/2}g_{\mu\nu}\delta \abs{F}^2 -
	\frac{1}{8}\mu_6 \frac{\sqrt{-g_3}}{\sqrt{-g_6}} \delta^3(\Sigma) g_{\mu\nu}e^{3\phi/4} \:.
\end{equation}
Again the result is zero and the eom is satisfied.

\section{$SU(3)$ Structure Conventions}

As said in the paper, the existence of the spinor $\eta$ implies the existence of a globally defined 2-form $J$ and 3-form $\Omega$:
\begin{gather}\label{eq005bis}
J_{mn} \equiv i \eta_{-}^\dagger \gamma_{mn}\eta_{-} = -i \eta_{+}^\dagger \gamma_{mn}\eta_{+} \\
\label{eq005b}
\Omega_{mnp} \equiv \eta_{-}^\dagger \gamma_{mnp}\eta_{+} \qquad
\Omega^*_{mnp} = -\eta_{+}^\dagger \gamma_{mnp}\eta_{-} \:,
\end{gather}
with the normalization $\eta_+^\dag \eta_+ = \eta_-^\dag \eta_- = 1$. $J$ and $\Omega$ satisfy:
\begin{gather} 
 {J_m}^n {J_n}^p = - \delta_m^p \\
 {(\Pi^+)_m}^n \Omega_{npq}=\Omega_{mpq}   \qquad   {(\Pi^-)_m}^n\Omega_{npq}=0 \\ 
 {(\Pi^\pm)_m}^n \equiv \frac{1}{2}(\delta_m^n \mp i{J_m}^n) \:.
\end{gather}
So $J$ defines an almost complex structures with respect to which $\Omega$ is $(3,0)$. Moreover
\begin{equation} \label{eq025} 
\Omega \wedge J=0 \qquad \mbox{and} \qquad J^3 = \frac{3i}{4} \Omega\wedge\Omega^* =  6 \dvol
\end{equation} 
and
\begin{equation}
\ast J = \frac{1}{2} J\wedge J  \qquad \ast (J \wedge J) = 2 J \qquad \ast \Omega = - i \Omega
\end{equation}
\begin{eqnarray}
\ast \tilde{F} = - \tilde{F}\wedge J && \ast (\tilde{F}\wedge J) = - \tilde{F}
\end{eqnarray}


\end{document}